\documentclass[%
 reprint,
superscriptaddress,
nofootinbib,
 amsmath,
 amssymb,
 aps,
 prx,
 longbibliography,
]{revtex4-2}

\usepackage{graphicx}
\usepackage{dcolumn}
\usepackage{bm}

\usepackage[a4paper,lmargin={2.2cm},rmargin={2.2cm},tmargin={2.2cm},bmargin={2.2cm}]{geometry}

\usepackage{amsfonts}
\usepackage{amsmath}
\usepackage{amssymb}
\usepackage{graphicx} 
\usepackage{caption}
\usepackage{subcaption} 
\usepackage{booktabs} 
\usepackage{tikz}
\usepackage{lipsum}
\usepackage{xargs} 
\usepackage[export]{adjustbox}  
\usepackage{placeins}
\usepackage{hyperref}
\usepackage{subfiles}
\usepackage{ragged2e}
\usepackage[colorinlistoftodos,prependcaption,textsize=tiny]{todonotes}

\definecolor{linkcolor}{HTML}{0b5394}

\hypersetup{
    colorlinks=true,
    linkcolor=linkcolor,     
    citecolor=linkcolor,    
    urlcolor=linkcolor       
}
\newcommand{\conrad}{C\textsc{onrad}}

\begin{document}

\title{Complete Optimal Non-Resonant Anomaly Detection}

\author{Gregor Kasieczka}
\email{gregor.kasieczka@uni-hamburg.de}
\affiliation{
    Institut f\"ur Experimentalphysik, Universität Hamburg \\
    Luruper Chaussee 149, 22761 Hamburg, Germany
}

\author{John Andrew Raine}
\email{john.raine@unige.ch}
\affiliation{
    Département de physique nucléaire et corpusculaire, University of Geneva\\
    1211 Geneva, Switzerland
}

\author{David Shih}
\email{shih@physics.rutgers.edu}
\affiliation{
New High Energy Theory Center, Rutgers University\\
    Piscataway, New Jersey 08854-8019, USA
}

\author{Aman Upadhyay}
\email{aman.u@rutgers.edu}
\affiliation{
New High Energy Theory Center, Rutgers University\\
    Piscataway, New Jersey 08854-8019, USA
}



\begin{abstract}

We propose the first-ever complete, model-agnostic search strategy based on the optimal anomaly score, for  new physics on the tails of distributions.
 Signal sensitivity is achieved via a classifier trained on auxiliary features in a weakly-supervised fashion, and backgrounds are predicted using the ABCD method in the classifier output and the primary tail feature.
The independence between the classifier output and the tail feature required for ABCD is achieved by first training a  
conditional normalizing flow that yields a decorrelated version of the auxiliary features; the classifier is then trained on these features. Both the signal sensitivity and background prediction require a sample of events accurately approximating the SM background; we assume this can be furnished by closely related control processes in the data or by accurate simulations, as is the case in countless conventional analyses. 
The viability of our approach is demonstrated for signatures consisting of (mono)jets and missing transverse energy, where the main SM background is $Z(\nu \nu) +\text{jets}$, and the data-driven control process is $\gamma+\text{jets}$.

\end{abstract}

 \maketitle

\section{Introduction}

The LHC has conducted numerous model-specific searches for new physics beyond the Standard Model, but thus far, none have yielded definitive proof of particles beyond the Standard Model. This has motivated an increasing interest in model-agnostic new physics searches in recent years, powered by modern machine learning, see e.g. \cite{Kasieczka:2021xcg, Karagiorgi:2022qnh, Darkmachines, Belis:2023mqs} for reviews and many original references.
A particularly promising approach has been to use data-driven techniques such as weakly-supervised classifiers 
and density estimation  
to approximate the Neyman-Pearson {\it optimal anomaly score}:
\begin{equation}\label{eq:optimal}
R(x)={p_{data}(x)\over p_{bg}(x)}
\end{equation}
and using this score to enhance the sensitivity to new physics in a model-agnostic way. 
Here, $x$ denotes some suitable input features while $p_{bg}$ and $p_{data}$ are the background and data densities, respectively. 
Much of the recent work has focused on learning the optimal anomaly score in the context of resonant new physics --- new physics localized in at least one kinematic feature (usually the invariant mass $M$ of some object in the 
event) --- where one can use sideband regions for data-driven estimates of the multi-dimensional background density ~\cite{Collins:2018epr,Collins:2019jip,anode,salad,Stein:2020rou,Amram:2020ykb,cathode,Collins:2021nxn,1815227,Kasieczka:2021tew,lacathode,Chen:2022suv,Kamenik:2022qxs,curtains, curtainsf4f,Golling:2023yjq, Bickendorf:2023nej, feta, Finke:2023ltw,Buhmann:2023acn,Sengupta:2023vtm,ranode}. 
Formally, this can be seen as interpolating $p_{bg}(x|M)$ from lower and higher values of $M$ into a signal region. Such 
resonant anomaly detection methods have also started to see their first applications to experimental data~\cite{ATLAS:2020iwa,CMS-PAS-EXO-22-026}.

However, learning $R(x)$ for new physics populating the tails of distributions --- as happens in many well-motivated scenarios of physics beyond the Standard Model (e.g. dark matter or supersymmetry~(SUSY)) --- is more difficult. The direct analogue of the resonant anomaly detection methods would be to {\it extrapolate} data-driven estimates of $p_{bg}(x|M)$ into the tail of $M$ (in this case $M$ might be a missing energy type variable), and the challenge is how to do this in a controlled way.

In this work, we present \conrad, the first-ever method for {\bf C}omplete, {\bf O}ptimal-anomaly-score-based  {\bf N}on-{\bf R}esonant {\bf A}nomaly {\bf Detection}. \conrad\ consists of a procedure for learning $R(x)$ on the tail of the $M$ distribution which is decorrelated from $M$ in the background. In this sense, \conrad\ is a complete strategy, in that it simultaneously enables model-agnostic signal detection via the optimal anomaly score, and accurate background estimation via the ABCD method.

\conrad\ leverages existing approaches to tail-based new physics searches. In a typical search, each SM background process has a set of associated reference samples -- either from simulation or, more often than not, from closely related control regions in the data. We assume such reference samples exist and use them to train a conditional normalizing flow to learn $p_{bg}(x|M)$. Then mapping the data events to the latent space of the flow, we can learn the anomaly score there and it is automatically statistically independent of $M$. Using conditional normalizing flows to decorrelate features has been shown to be effective for both resonant anomaly detection~\cite{lacathode} and fully supervised classification tasks~\cite{Klein:2022hdv}.
In this work, we show that it works equally well for non-resonant anomaly detection.

In this proof-of-concept work, we illustrate our ideas for non-resonant anomaly detection using the well-studied
(mono)jet+MET final state \cite{ATLAS:2015qlt,ATLAS:2016bek,CMS:2017jdm,ATLAS:2017bfj,CMS:2017zts,ATLAS:2021kxv,CMS:2021far} where the main SM background is from $Z(\nu\nu)+\mathrm{jets}$.
A closely-related control process is  provided by $\gamma+\mathrm{jets}$, where the momentum of the photon can be used in place of MET. In this first study, we restrict our attention to the 
case where the control process is fully data-driven.
In future work we aim to extend this setup to the simulation assisted case, where  additional samples could be used to provide corrections and refinements to further improve the signal sensitivity and accuracy of the background modelling.

The outline of our paper is as follows: in Section~\ref{sec:methods}, we describe alternative attempts for non-resonant anomaly detection and then discuss \conrad\ in more detail. Section~\ref{sec:setup} contains the setup of our proof-of-concept demonstration. Results are described in Section~\ref{sec:results}, and we conclude in Section~\ref{sec:conclusions}.  

\section{Methods}

\label{sec:methods}

\subsection{Related works}

In \cite{Mikuni:2021nwn}, it was shown how two decorrelated autoencoders could be trained to find non-resonant anomalies in conjunction with the ABCD method for background estimation. While this was a complete strategy for tail-based anomaly detection, it was not based on the optimal anomaly score (\ref{eq:optimal}).
Recently \cite{Bickendorf:2023nej} emphasized how methods for resonant anomaly detection such as CATHODE \cite{cathode} could also be applied to find new physics where some features are resonant and other features are on tails. 
In~\cite{DAgnolo:2018cun,dAgnolo:2021aun,Letizia:2022xbe}, it was demonstrated how a reliable test statistic for new physics --- sensitive in principle to phenomana in the bulk or tail of a distribution --- can be constructed. However, this construction already assumes knowledge $p_{bg}(x)$ while we demonstrate how it can be adopted from data.

Finally, \cite{Bai:2023yyy} developed methods for extrapolating and refining backgrounds to the tails of distributions, also motivated by model-agnostic searches for non-resonant new physics with the optimal anomaly score.
There, the signal was considered to be on the tail of a {\it multi}-dimensional space $M$ (\cite{Bai:2023yyy}  focused on $M\in {\Bbb R}^2$).
The control region was defined by inverting the signal region selection and used to learn an approximation of $p_{bg}(x|M)$, which was then extrapolated into the signal region (assisted by simulation) and used to train an optimal classifier $R(x)$.  
However, these methods were incomplete in that they did not attempt to control for the evolution of $p_{bg}(x|M)$ onto the tails of the joint $M$ distribution. Furthermore, they did not demonstrate closure of a background estimation method after a cut on the learned anomaly score, a necessary ingredient to obtain statistically robust results. In this respect, it would be interesting to explore combinations of \conrad\ with the methods proposed in  \cite{Bai:2023yyy}, to potentially cover final states where reliable background templates are not readily available.

\subsection{CONRAD}
\label{sec:method}

Our goal is to enhance tail-based searches for non-resonant new physics using model-agnostic machine learning. We imagine starting with an {\it inclusive signal region}, which is defined in terms of some kinematic feature $M$ (e.g. the missing transverse energy) to be
\begin{equation}
M > M_c
\end{equation}
This is the analogue of the mass window for resonant anomaly detection. Let $x$ be an additional set of features (e.g.\ $H_T$, number of jets, jet $p_T$ and substructure variables, etc).
We aim to learn an anomaly score $R(x)$ that enhances the sensitivity to signal in this inclusive signal region. As in other overdensity detection methods, we will learn $R(x)$ using a classifier between data in the inclusive signal region and a events drawn from a data-driven background model.

The final, anomaly-enhanced signal region is then defined by a combination of cuts on $M$ and $R(x)$:
\begin{equation}
 M>M_c\,\,\,{\rm and}\,\,\,R(x)>R_c.
\end{equation}
The threshold $R_c$ can be optimized in the final stage of the analysis, or (as is often the case) chosen in a model-agnostic way using a fixed working point (e.g.\ chosen to keep 1\% or 0.1\% of the data).

\begin{figure}[t]
    \centering
    \includegraphics[width=\linewidth]{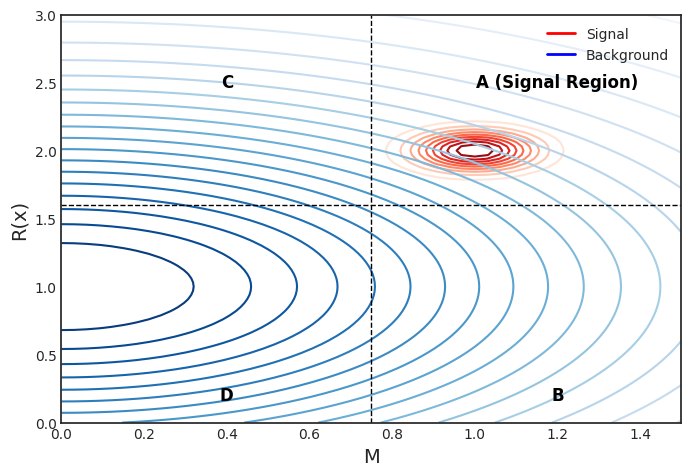}
    \caption{Representation of the four regions of the ABCD method defined by the cuts on $R(x)$ and $M$ (in arbitrary units), where A is the signal region, with example signal (red) and background (blue) processes.}
    \label{fig:ABCD}
\end{figure}

In any viable new physics search strategy, one requires an accurate estimation of SM backgrounds in addition to signal sensitivity. Here, background estimation will be achieved through the ABCD method between $R(x)$ and $M$. 
This requires $R(x)$ and $M$ to be independent for the background. In that case, the background estimate in the signal region can be easily calculated from the background yields  $N_{bg}$ in the three control regions defined by inverting these cuts. 
Labelling the regions as A, B, C and D, as in Fig.~\ref{fig:ABCD}, the background yields in the signal region are given by
\begin{equation}
    N_{A,bg}^{pred}=\frac{N_{B,bg}\cdot N_{C,bg}}{N_{D,bg}}.
\end{equation}

\begin{figure*}[t]
    \centering
    \includegraphics[width=0.4\linewidth]{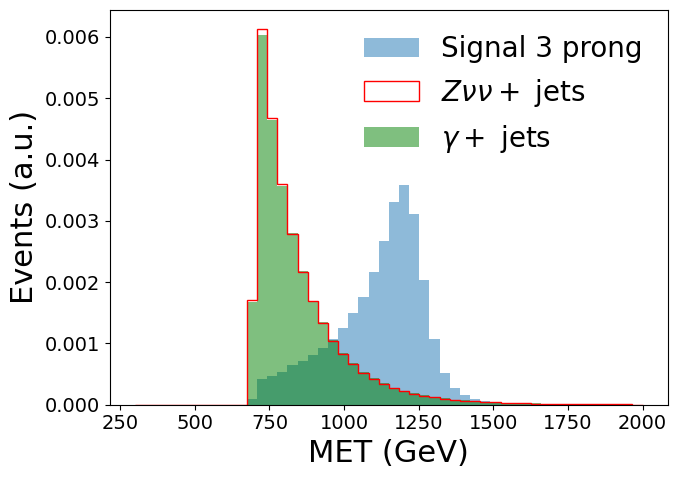}
    \includegraphics[width=0.4\linewidth]{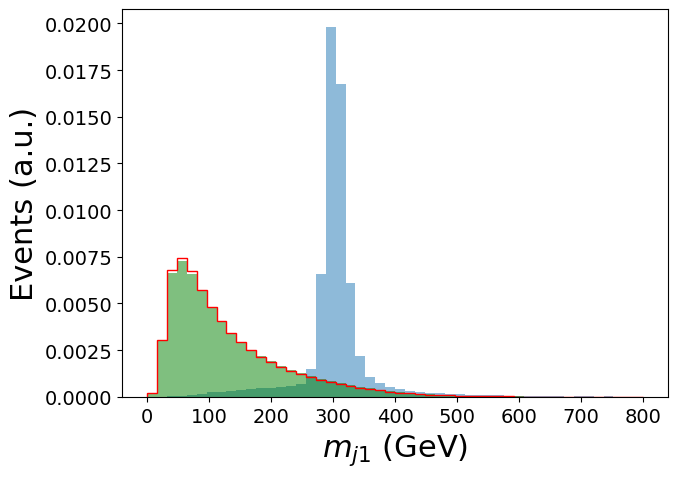}
    \includegraphics[width=0.4\linewidth]{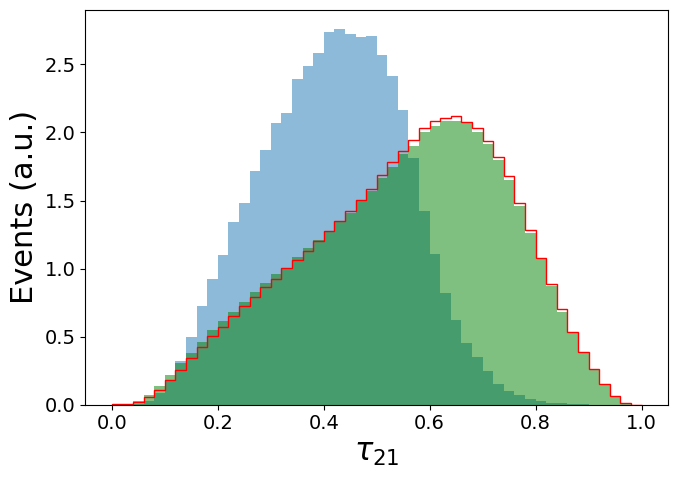}
    \includegraphics[width=0.4\linewidth]{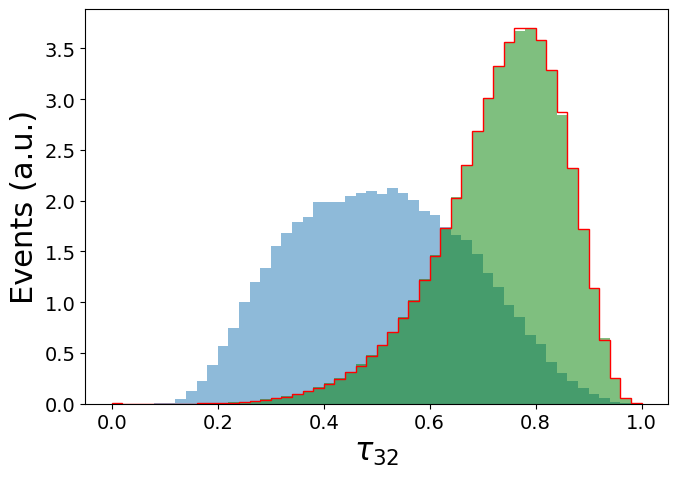}
    \caption{Normalized distributions of features.}
    \label{fig:19}
\end{figure*}

To achieve the required statistical independence of $R(x)$ and $M$ in the background, we propose to train a conditional normalizing flow~\cite{NormFlows1,NormFlows2} on background events to learn $p_{bg}(x|M)$. The flow represents a bijective (in $x$) transformation from pairs $(x,M)$ to a latent space~$z$, and training a classifier between data and background in $z$-space results in an anomaly score which is automatically decorrelated from $M$ \cite{lacathode,Klein:2022hdv}.  

As described in the Introduction, we assume that a sample of background-only events can be furnished through a combination of data-driven control regions (consisting of closely related background processes) and simulations. This is the case in countless conventional LHC analyses, and one could view \conrad\ as a general strategy to enhance the model-independence of these analyses.

\section{Setup}
\label{sec:setup}

For the proof-of-concept jets+MET search studied in this work, we consider $Z(\nu\nu)+$jets to be the actual background.\footnote{In reality, the jets+MET final state also receives significant contributions from other SM processes such as $W(\ell\nu)+$jets and $t\bar t$, but we are ignoring these additional processes for simplicity for this initial proof-of-concept study.} At high $p_T$, $\gamma+$jets events are sufficiently similar to $Z(\nu\nu)+$jets \cite{Ask:2011xf} that they can be used as an accurate control sample once the identified photon is ``deleted" from the event and its $p_T$ included in the definition of the MET, see e.g.~\cite{CMS:2009wxa,CMS:2011xek,CMS:2011yke,ATLAS:2011hnu,ATLAS:2011kfa} for some early LHC examples of this technique.

\subsection{Datasets}

Data for this paper was generated from $pp$ collisions at 13~TeV using \textsc{Pythia}~8.219~\cite{Pythia} (with default tune) interfaced with \textsc{Delphes}~3.4.1~\cite{Delphes} using the default CMS card. $R=1$ jets were clustered with \textsc{FastJet}~3.3.0~\cite{FastJet} using the anti-$k_t$ algorithm~\cite{AntiKt}.

For SM background processes, we generated $Z(\nu\nu)+$jets and $\gamma+$jets using \textsc{Pythia} built-in $2\to 2$ processes; any additional jets were produced using default \textsc{Pythia} ISR settings.

For the signal model, the following benchmark was used for this proof-of-concept study: $Z'\to XY$, with $X\to qqq$ and $Y\to{\rm invisible}$, with $m_{Z'}= 2.5$~TeV, $m_{X}= 300$~GeV, and $m_Y=100$~GeV.\footnote{We note that this choice of signal process does have a resonance at $m_X=300$~GeV (see Fig. \ref{fig:19})  so it is possible that enhanced bump hunt methods~\cite{anode,salad,cathode,curtains,feta,curtainsf4f}
could also find this signal effectively (despite the presence of non-resonant features~\cite{Bickendorf:2023nej}), but here we use these signals to demonstrate the effectiveness of non-resonant anomaly detection.} 
This is a variation on one of the LHCO R\&D signal datasets~\cite{LHCOlympics}, where originally $Y$ also decayed visibly. Thus every signal event has a high-$p_T$ large-radius mono-jet with non-QCD-like jet substructure and significant missing energy. In appendix~\ref{app:twoprong}, we also consider two-prong decays of $X$, where our method  gives similar performance.

As input observables, we take the protected tail feature to be
\begin{equation}
M = \mathrm{MET}
\end{equation}
and the additional features $x$ to be three properties of the highest $p_T$ large radius jet, namely its invariant mass $m_J$, and the two $N$-subjettiness ratios~\cite{nsubjettiness} $\tau_{21}$ and $\tau_{32}$,
\begin{equation}\label{eq:xfeatures}
x=(m_{J_1},\tau_{21}^{J_1},\tau_{32}^{J_1})
\end{equation}
The distributions of the features for the background and signal processes are shown in Fig.~\ref{fig:19}. 

We emulate the effects of a trigger by requiring $p_T(j_1)>500$~GeV and $\text{MET}>700$~GeV for all events.
In addition, we will define the inclusive signal region to be the tail region
\begin{equation}
\mathrm{MET}>1000~{\rm GeV}
\end{equation}

After the trigger, our jets+MET data consists of 700k $Z(\nu\nu)+$jets events, of which 90k are in the inclusive signal region, mixed with various levels of anomalous signal events.
For an initial 3$\sigma$ significance in the inclusive signal region, this corresponds to 1200 signal events of which 947 are in the inclusive signal region.

We will consider two versions of the analysis: a ``fully-idealized" version with a background template consisting of $Z(\nu\nu)+$jets events drawn from the same generator as those in the jets+MET data; and a more realistic version where the background template consists of $\gamma+$jets events.

For the fully-idealized (more realistic) case, 776k $Z(\nu\nu)+$jets (956k $\gamma+$jets) events are used for training the conditional normalizing flow. 
This flow is used to map jets+MET data into the latent space, where a classifier is trained to distinguish it from background events drawn from the normal distribution.\footnote{In principle, if the flow is not perfectly normalizing the background events, one could get a different result if classifying data vs.\ background events mapped to the latent space. We have checked that this is not an issue here. We have also checked that oversampling the background \cite{cathode} does not provide any benefit here.}

In a real analysis, all of the jets+MET data would be utilized in the final step, through $k$-fold cross validation. In this initial proof-of-concept study, we skip this cumbersome step for simplicity and just divide the jets+MET data into two equal halves. The first half will be used for training the classifier, which will then be applied to the second half to evaluate the performance of the method. This way each half of the jets+MET data in the inclusive signal region has $\frac{S}{\sqrt{B}} \approx 2$.  The testing dataset with 2$\sigma$ will be used for evaluating the method's effectiveness in enhancing significance.

\subsection{ML architectures}

We construct the conditional normalizing flow using masked autoregressive flows with rational quadratic spline~(RQS) transformations~\cite{NeuralSplines}, using the \texttt{nflows} library~\cite{NFlows} with \texttt{pytorch}~\cite{pytorch}.
The model comprises four RQS spline transformations \cite{NeuralSplines} each with 8 knots. Input data are transformed to a latent space defined by a standard normal distribution.
We train the model for 50 epochs with the Adam optimizer with a learning rate of $10^{-4}$ and a batch size of 128.
An ensemble of five normalizing flows are trained to ensure stability of the latent space.
The normalizing flows are trained inclusively of the event selection, in order to ensure there is no extrapolation.

For our classifier we employ histogram gradient boosting trees as implemented in the \texttt{sci-kit learn} package~\cite{scikit-learn}, due to their observed robustness to irrelevant features and improved performance over feed-forward neural networks~\cite{Finke:2023ltw,Freytsis:2023cjr}.
We use an ensemble-average of 50 boosted decision trees as our classifier. Each boosted decision tree is trained with a maximum iteration of 200 and maximum leaf nodes of 61. The data kept aside to train the classifier is then split again and $75\%$ of the data is used for training the tree, and $25\%$ is kept for classifier validation.  Similarly, we train another classifier with 50 BDTs for our data-driven background template using the  $\gamma +$jets as our background template.

\section{Results}
\label{sec:results}

\subsection{Signal sensitivity}

\begin{figure}[b]
    \centering
    \includegraphics[width=\linewidth]{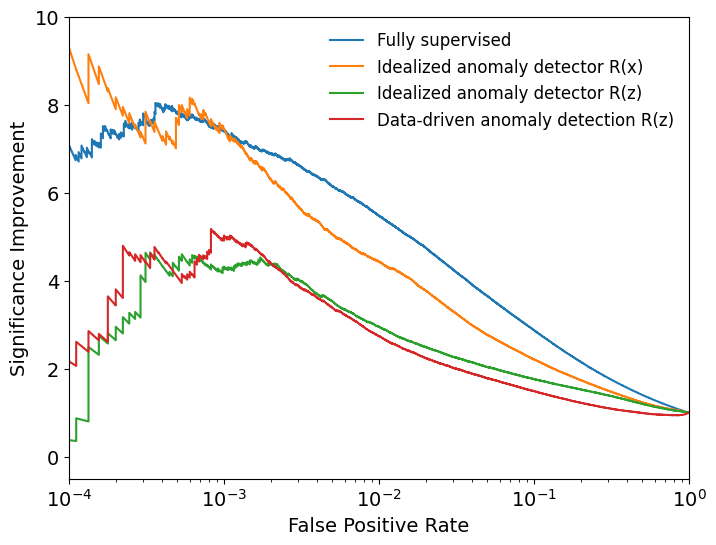}
    \caption{Significance improvement vs False Positive Rate (FPR) for different classifiers for the three-prong signal.}
    \label{fig:sic_curves_threeprong}
\end{figure}

Shown in Fig.~\ref{fig:sic_curves_threeprong} are the significance improvement characteristic (SIC) curves for the three-prong signal. Here SIC is defined as
\begin{equation}
{\rm SIC}={\epsilon_S\over\sqrt{\epsilon_B}}
\end{equation}
where $\epsilon_{S,B}$ are the signal and background efficiencies after a cut on the classifier. These efficiencies are measured using an additional 45k $Z(\nu\nu)+$jets and 60k signal events not used in the training of the flow or the classifier.

The blue curve corresponds to a fully supervised signal vs.\ background classifier trained on the $x$ features from Eq.~(\ref{eq:xfeatures}) in the inclusive signal region. This is meant to represent an upper limit on the best possible significance improvement for this signal given the features chosen for the study. We see that the SIC can reach as high as a factor of 8, which would represent an sizable enhancement of the 2$\sigma$ signal significance in the testing dataset.

Its weakly-supervised counterpart -- the idealized anomaly detector trained between jets+MET and perfectly simulated $Z(\nu\nu)+$jets background -- is shown in orange; any difference between these two should be ascribed to limited training statistics, noisy or uninformative features, and finite model capacity. In theory, asymptotically and with the most optimal classifier, there should be no difference between the blue and orange curves. 

Neither of these could be combined with the ABCD method for background estimation, since the features $x$ are still correlated with MET. Shown in green is the result of training the idealized anomaly detector in latent space $z$ that has been decorrelated against MET. We see that the performance drops to maximum SIC below 5. This probably has a variety of causes, including the fact that the original $x$ features are correlated with MET, so the classifier in $x$ space was also leveraging the discriminating power of MET, but this is not available in the decorrelated $z$ space. Another possible cause of the performance drop is that the BDT is less optimal in the latent space than in the original data space, i.e.\ the latent space is somehow ``less tabular" than the data space. 

Finally, the more realistic, data-driven case --- i.e. utilizing the full \conrad\ protocol --- is shown in red: a classifier trained on jets+MET data vs.\ a background template drawn from $\gamma+$jets data, in the latent space derived from the latter. Here we see basically no performance drop from the idealized case. This is strong evidence that $\gamma+$jets is a good proxy for $Z(\nu\nu)+$jets in these features. Overall, we conclude that the fully data-driven non-resonant anomaly detector can nominally enhance the significance of the signal by up to a factor of $4-5$, which would raise it well above discovery significance if it were fully realized.

\subsection{Background estimation}

Having explored the signal sensitivity of the weakly-supervised anomaly score, we now investigate the accuracy of the background estimation by the ABCD method. 

\begin{figure}[t]
    \centering
    \includegraphics[width=\linewidth]{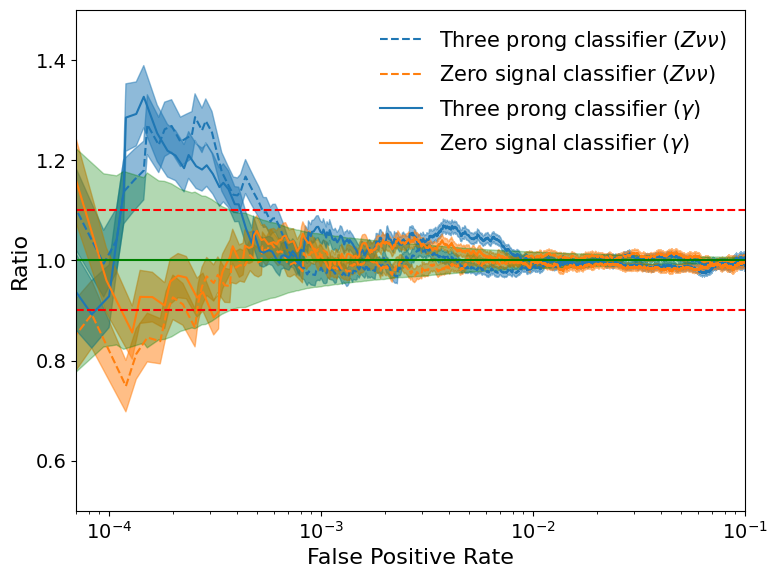}
    \caption{Ratio of ABCD background estimation and the true background vs. false positive rate.}
    \label{fig:18}
\end{figure}

Starting with the idealized case where $Z(\nu\nu)+$jets is used as the background template, we show in Fig.~\ref{fig:18} the ratio of predicted (from the ABCD method) to the true number of backgrounds,
\begin{equation}\label{eq:ratio}
r = {N_{A,bg}^{pred}\over N_{A,bg}}
\end{equation}
as a function of the false positive rate of the classifier. The error bands on the ratio are obtained by propagating the Poisson-derived uncertainty on $N_{A,bg}^{pred}$ through the ratio (\ref{eq:ratio}), while the error bands on $r=1$ correspond to the propagation of the Poisson uncertainty on $N_{A,bg}$ through the ratio. The blue curves correspond to the benchmark three-prong signal while the orange curves illustrates the amount of closure in the absence of any signal. We see from Fig.~\ref{fig:18} that the accuracy of the background prediction of \conrad\ is fairly stable down to a false positive rate of $\sim 10^{-3}$. It generally remains accurate to within 10\%, and stays within the envelope set by the Poisson uncertainty of $N_{A,bg}$, whether signal is present or not. This indicates that we have succeeded in achieving a high degree of decorrelation against MET while retaining sensitivity to the signals.

Meanwhile, Fig.~\ref{fig:18} also shows the same ratio plots for the more realistic $\gamma+$jets background template.  (Here although $N_{A,bg}^{pred}$ in Eq.~\ref{eq:ratio} is derived from the $\gamma+$jets control sample, $N_{A,bg}$ is still taken from the $Z(\nu\nu)+$jets in the jets+MET data.) Encouragingly, we see that the background estimation is accurate and in agreement with $Z(\nu\nu)+$jets down to low false positive rates also in this case.

\begin{figure}[t]
    \centering
    \includegraphics[width=\linewidth]{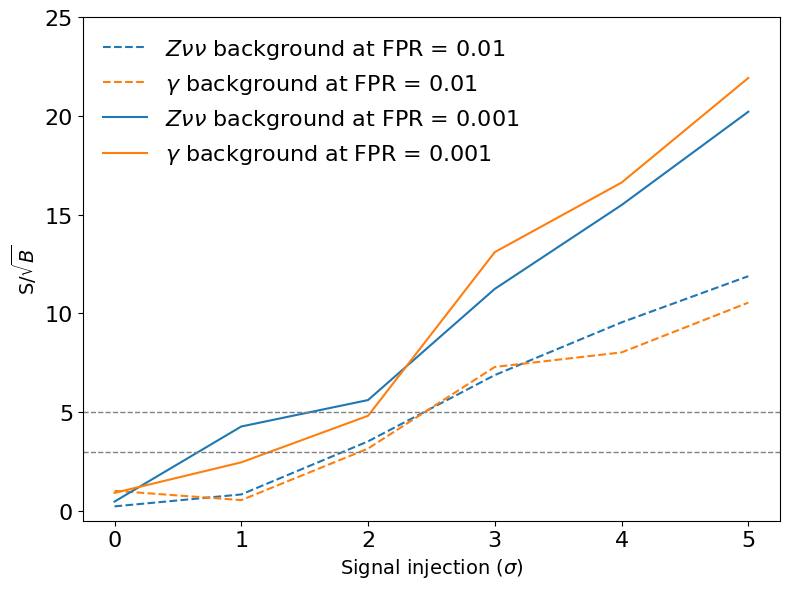}
    \caption{$\frac{S}{\sqrt{B}}$ vs. signal injection.}
    \label{fig:signal_senitivity}
\end{figure}

Finally, in Fig.~\ref{fig:signal_senitivity}, we combine the signal sensitivity and background estimation and show the actual $S/\sqrt{B}$ significance expected from the search, where now $B=N_{A,bg}^{pred}$ is the number of ABCD predicted background events, and $S=N_{A,obs}-N_{A,bg}^{pred}$ is calculated as the difference of the number of events in the signal region and predicted background $B$. We show this as a function of the amount of signal injected (the initial significance in the inclusive signal region), for two different working points corresponding to 1\% and 0.1\% false positive rates. We see good agreement between using the true $Z(\nu\nu)$+jets background to derive the anomaly score and using the $\gamma$+jets proxy. We also see that the expected significance drops as the amount of signal drops, as is to be expected in an approach based on weak supervision. The fact that it falls to zero at zero signal is reassuring, as it indicates proper closure and lack of false positives in the absence of signal. Finally, we see the overall performance is very promising, as expected from the previous plots of the SIC curves and ABCD prediction ratio -- at the tighter working point, we can turn a 2$\sigma$ excess in the signal region into a 5$\sigma$ discovery, using our proposed non-resonant anomaly detection method.

\section{Conclusions}
\label{sec:conclusions}

In this work we have presented \conrad, the first complete non-resonant anomaly detection method for BSM physics searches based on the optimal anomaly score, taking into account both signal sensitivity and accurate background modelling.
The approach is completely data driven, and 
is able to turn a $2\sigma$ excess in the tail into over a $5\sigma$ discovery for the benchmark three-prong signal process considered.

Although some deviation from the true background rate is observed with the ABCD method, this is within the statistical uncertainties of the true background contribution. 
As a future development, we propose to improve the agreement between the $\gamma+$jets control sample and the $Z(\nu\nu)+$jets background with data driven reweighting or transport functions similar to those studied for correcting for mis-modelling in MC simulation~\cite{salad,feta,Golling:2023mqx}.
We expect this should further improve the performance of the method, and would also be necessary for applications where no control sample in the data is available.

\bigskip

\section*{Acknowledgments}

We are grateful to B.~Nachman for helpful discussions and feedback on the draft.
The work of AU and DS was supported
by DOE grant DE-SC0010008. GK acknowledges support by the Deutsche Forschungsgemeinschaft under Germany’s Excellence Strategy – EXC 2121  Quantum Universe – 390833306.
JR is supported by the SNSF Sinergia grant CRSII5\_193716 ``Robust Deep Density Models for High-Energy Particle Physics and Solar Flare Analysis (RODEM)''
and the SNSF project grant 200020\_212127 ``At the two upgrade frontiers: machine learning and the ITk Pixel detector''.

\begin{figure*}[t]
    \centering
    \includegraphics[width=0.8\linewidth]{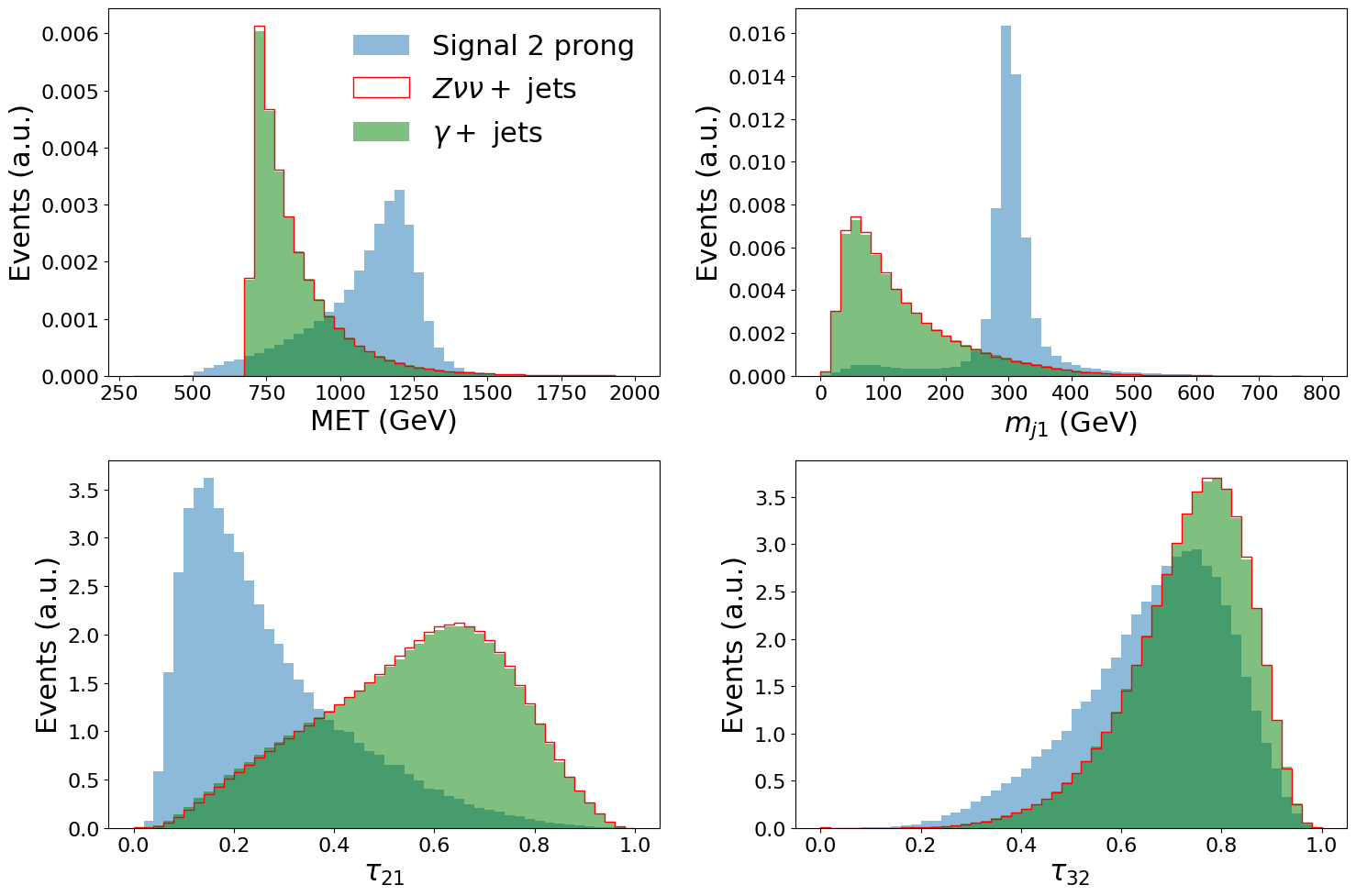}
    \caption{Normalized distribution of features with  two-prong signal.}
    \label{fig:20}
\end{figure*}

\appendix

\section{Two-prong signal}
\label{app:twoprong}

While the body of the paper focused on the demonstration of \conrad\ for the case of a three-pronged signal, a key motivation for carrying out searches for new physics via detecting anomalies, is sensitivity to a broad range of potential signals. Here, we show that the approach works equally well for a two-pronged signal. The signal model considered here was  $Z'\to XY$, with $X\to qq$ and $Y\to{\rm inv}$, with $m_{Z'}= 2.5$~TeV, $m_{X}= 300$~GeV, and $m_Y=100$~GeV. 1200 signal events (931 of which are in the inclusive signal region) are again mixed with the same number of $Z(\nu\nu)+$jets events. This data was again divided into two equal halves, the first used to train the classifier and the second used to evaluate the performance of the method. Each half of the jets+MET data in the inclusive signal region has $\frac{S}{\sqrt{B}} \approx 2$.

First, Fig.~\ref{fig:20} shows the differential distributions of input features. As expected, the difference in substructure now is now marked in the $\tau_{21}$ n-subjettiness ratio, while $\tau_{32}$ becomes more background like.

\begin{figure}[t]
    \centering
    \includegraphics[width=\linewidth]{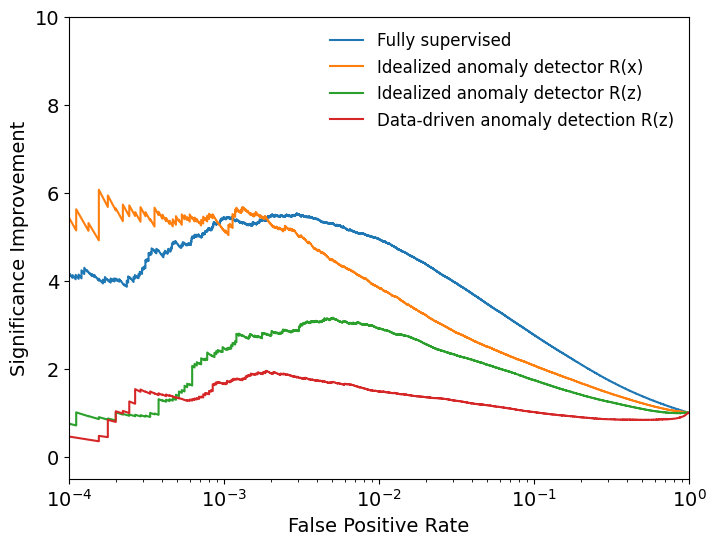}
    \caption{Significance improvement vs False Positive Rate (FPR) for different classifiers for the two-prong signal.}
    \label{fig:sic_curves_twoprong}
\end{figure}

\begin{figure}[t]
    \centering
    \includegraphics[width=\linewidth]{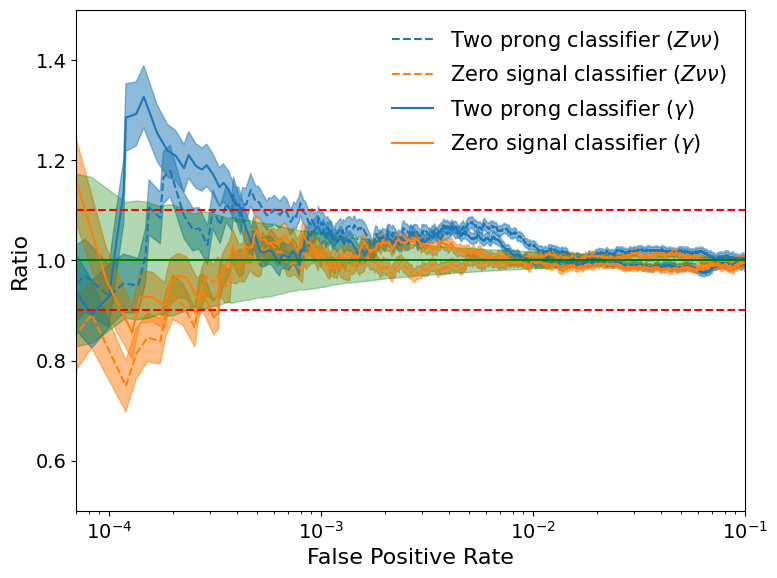}
    \caption{Ratio of ABCD background estimation and the true background vs. false positive rate with two prong signal.}
    \label{fig:18_1_2prong}
\end{figure}

\begin{figure}[t]
    \centering
    \includegraphics[width=\linewidth]{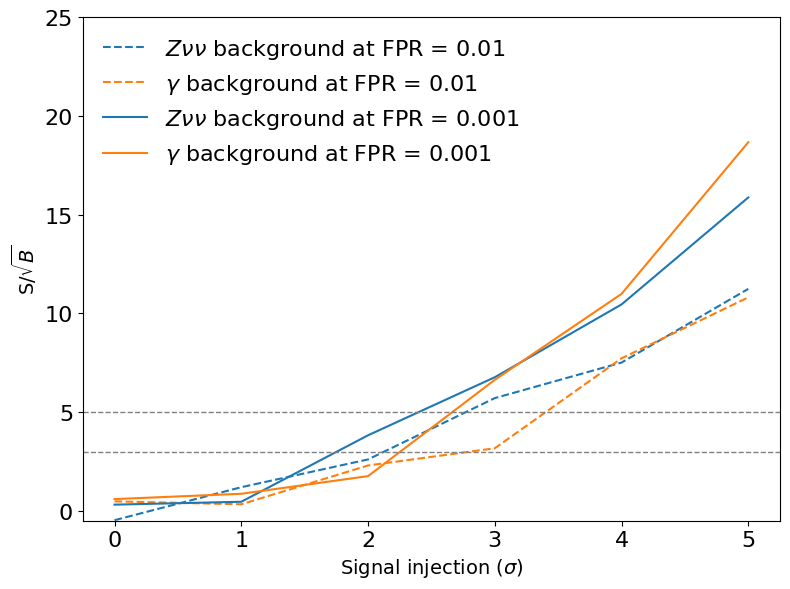}
    \caption{$\frac{S}{\sqrt{B}}$ vs. signal injection with two prong signal.}
    \label{fig:19_1_2prong}
\end{figure}

Next, Fig.~\ref{fig:sic_curves_twoprong} shows the corresponding SIC curves. As for the case of the three-prong signal, the fully-supervised approach and the idealised anomaly detector in $R(x)$ achieve a similar performance with a maximum SIC value around 6. We observe a similar reduction in performance when moving to the decorrelated latent space $R(z)$ with a maximum SIC around 3. Finally, the upshot from the red curve is a modest significance improvement of the two-prong signal, reaching a maximum SIC of $\sim 2$. 
On the other hand, the background estimation in Fig.~\ref{fig:18_1_2prong} remains stable and \conrad\ achieves 5 $\sigma$ sensitivity for 2.4 and 2.6 $\sigma$ significance in the testing dataset using the $Z\nu \nu$ and $\gamma$-derived backgrounds respectively according to Fig.~\ref{fig:19_1_2prong}.

\bibliography{main}

\end{document}